\documentclass[twocolumn,showpacs,preprintnumbers]{revtex4}
\usepackage{amsmath}
\usepackage{amssymb}
\usepackage{graphicx}

\begin{document}
\title{Strongly correlated  quantum dots in weak confinement potentials and magnetic fields}
\draft
\author{Min-Chul~Cha$^{a}$ and S.-R. Eric Yang$^{b}$}
\address{$^a$ Department of Physics, Hanyang University, Ansan 425-791,Korea\\
 $^b$ Department of Physics, Korea University, Seoul, Korea }
\date{\today}

\begin{abstract}
We explore a strongly correlated quantum dot in
the presence of a weak confinement potential and a weak magnetic field. 
Our exact diagonalization studies show that the ground state property of such a quantum dot 
is rather sensitive to the magnetic field and the strength of the
confinement potential.  
We have determined rich phase diagrams of these quantum dots.
Some 
experimental consequences of the obtained phase diagrams are discussed.

\end{abstract}
\maketitle

Fabrication of quantum dots has reached such an advanced state that 
the shape of the confinement 
potential and the number of electrons in them can be tuned precisely \cite{review}.
Ground states of isolated  quantum dots fabricated from 2D electron systems 
can  display various  strong correlation effects. 
Quantum Hall related ground states have been explored in the strong magnetic field limit \cite{yang,lll},
and 
Hund's rule for shell structures for a two-dimensional harmonic potential has been investigated \cite{hund}.
Recently transport through a quantum dot has attracted  much attention, especially
Kondo related physics \cite{kondo}. 
Wigner crystal states and strong electron correlation in quantum dots  have been explored
through various methods\cite{papers,mik}

In this paper we explore possible ground states of an isolated quantum dot made of dilute 2D electron systems 
in the presence of a weak magnetic field.  Recently unusual behavior suggestive of a metal-insulator 
transition has been reported in a variety of dilute two-dimensional electron and hole systems \cite{abr}, where
the dimensionless parameter $R=E_{e-e}/E_s \gg 1$ (here $E_{e-e}$ is the electron-electron interaction energy
and $E_s$ is the characteristic single particle energy). Strong electron correlation effects are expected 
to be play an important role in these systems. 
Quantum dots fabricated from dilute 2D electron systems are also expected to exhibit  
strong electron correlation effects.  Here  we focus on how the ground state spin depends
on the applied magnetic field and the strength of the confinement potential.
To describe properly the delicate competition\cite{pfa} between
different spins states in strongly correlated regime 
non-perturbative methods\cite{exact,yang,pal,silj} are required. 
We adopt exact diagonalization methods\cite{exact,yang,pal}
since various mean field theoretical methods\cite{mean}
are applicable only in the regime $R < 1$.

We model a quantum dot as follows.
Electrons move on a 2D plane under the influence of 
a parabolic confinement potential and a weak magnetic field
applied perpendicular to the plane.
In parabolic quantum dots, the ratio $R$ can be characterized
by $(e^2/\kappa a)/\hbar\Omega$ where $a$ is the typical lateral size
of the dot, $\kappa$ the dielectric constant,
and $\Omega$ the confining frequency of the harmonic potential.
We consider up to six electrons in the parameter regime
$R \sim 5-16$.
We find that the ground state of quantum dots for these values of $R$ 
is rather sensitive to the magnetic field and the strength of the 
confinement potential. 
Our investigation suggests that this is a direct consequence of
strong electron correlation: it originates from the existence of
nearly degenerate quantum dot eigenstates.
We obtain rich phase diagrams of these quantum dots and discuss  
experimental consequences.

Electrons of the dot is confined by 
a harmonic potential $V(r)={ 1 \over 2} m^* \Omega^2 r^2$,
where $m^*$ is the effective electron mass.
A magnetic field $B$ is applied along the $z$-axis
through a  vector potential in a symmetric gauge. We include also
the Zeeman splitting, $g\mu B [{\rm meV}]=0.026B[{\rm T}]$.
We take the  Hamiltonian of $N$ electron  dot  
to be  $H=\sum_{i=1}^{N}H_i+H_{int}$,
where the single particle Hamiltonian is modeled by 
\begin{eqnarray} 
H_i = {{\bf p}_i^2 \over 2m^*} + { 1 \over 2} m^* \omega^2 r_i^2
-{1 \over 2}  \omega_c {\bf l}_i \cdot {\bf \hat B}-g\mu  B S_{z,i},
\end{eqnarray}
and where the many-body interaction term is given by
\begin{eqnarray}
H_{int}= {1 \over 2}\sum_{i \ne j}^N
{e^2 \over \kappa |{\bf r_i} - {\bf r_j}|}.
\end{eqnarray}
Here $\kappa=12.4$ and $\omega_c $ are the dielectric constant of GaAs 
semiconductor and cyclotron frequency.
The eigenfunctions\cite{fock,darw} of the single particle Hamiltonian are 
\begin{eqnarray} 
\phi_{nls_z}({\bf r})={1 \over \sqrt{2\pi} a} e^{-il\theta}
R_{nl}({r^2 \over 2 a^2})\chi_{s_z}
\end{eqnarray}
where  $a^2=\hbar/(2m\omega)$, $\omega^2=\Omega^2+{1\over 4}\omega_c^2$, $\chi_{s_z}$ 
are spin functions, and
\begin{eqnarray} 
R_{nl}(x)=\sqrt{n! \over (n+|l|)!} e^{-x/2} x^{|l| \over2} L_n^{|l|}(x).
\end{eqnarray}
Note that, as expected, the dependence on $\theta$ indicates that
electrons rotate clockwise when a magnetic field applied along the $z$-direction.  
However, it is convenient to
define the  $z$-component of single particle angular momentum as 
${\bf l}_z=-\frac{1}{i}\frac{\partial}{\partial\theta}$ so that 
the eigenstates with positive angular momenta have lower energy.
The eigenenergies are
\begin{eqnarray}
\epsilon_{n l s_z} = \hbar \omega (2n+|l|+1)-\frac{1}{2}\hbar\omega_c\ell-g\mu
S_z B.
\end{eqnarray}
These eigenstate wavefunctions
are labeled by quantum numbers of orbital states
and the $z$-component of angular momentum $(n,l)$:
Note that for a given $n=0,1,2,...$ the quantum number $l$ can take
all possible integer values.
It is useful to study  properties of {\it zero} magnetic field single particle states.
The eigenenergies are
\begin{eqnarray}
\epsilon^0_{n l} = \hbar \Omega (2n+|l|+1).
\end{eqnarray}
These single particle energy levels $\epsilon_{nl}^{0}$ can be degenerate,
for example, $(n,l)$ and $(n,-l)$.
These degeneracies are such that they lead to 
the magic numbers of harmonic potential 2, 6, 12, etc.
In the presence of a weak magnetic field along the $z$-axis {\it all} degeneracies will be lifted.

The many-body ground state $\Psi$ of $H$ is expanded in terms of  Slater determinant states 
\begin{eqnarray}
\Psi=\sum_{\alpha}C_{\alpha}\Psi_{\alpha},
\end{eqnarray}
where
\begin{eqnarray}
\Psi_{\alpha}&=& c_{n_1 l_1 \sigma_1}^{+}c_{n_2 l_2 \sigma_2}^{+}...
c_{n_N l_N \sigma_N}^{+}\Psi_0.
\end{eqnarray}
Here $\alpha=\{n_1 l_1 \sigma_1,...,n_N l_N \sigma_N\}$ and labels a Slater determinant.
Each creation operator $c_{n ls_z}^{+}$ creates an electron in an eigenstate of
{\it zero} magnetic field single particle Hamiltonian.
The energy of $\Psi_{\alpha}$ is
denoted by $E_{\alpha}$, and it is the sum of energies of single particle states
$\phi_{nls_z}$ 
Even in the presence of a magnetic field we use the same many-body basis states
as in zero magnetic field, with replacing $\Omega$ by $\omega$.
We found numerically that it is convenient to use these zero field 
particle states in constructing Slater determinants.
The many-body Hilbert space can be divided into
regions with different quantum numbers $(S_z,L_z)$:
In each region we construct the many-body basis states
which have the well-defined $z$-component of total angular
momentum quantum number $L_z=\sum_{i=1}^N l_i$ and
the $z$-component of total spin  $S_z$.
All possible eigenstates in the Hilbert space $(S_z,L_z)$
are calculated, and the minimum energy state is found.

\begin{figure}
\center
\includegraphics*[width=2.5in,height=2.5in]{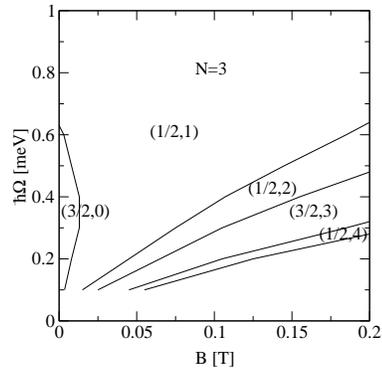}
\caption{Phase diagram for  $N=3$.
Each state is denoted by the quantum numbers $(S_z,L_z)$.
For strongly correlated regime (for small $\hbar\Omega$),
the ground states are very sensitive to the magnetic field.}
\label{fig:fig1}
\end{figure}

\begin{figure}
\center
\includegraphics*[width=2.5in,height=2.5in]{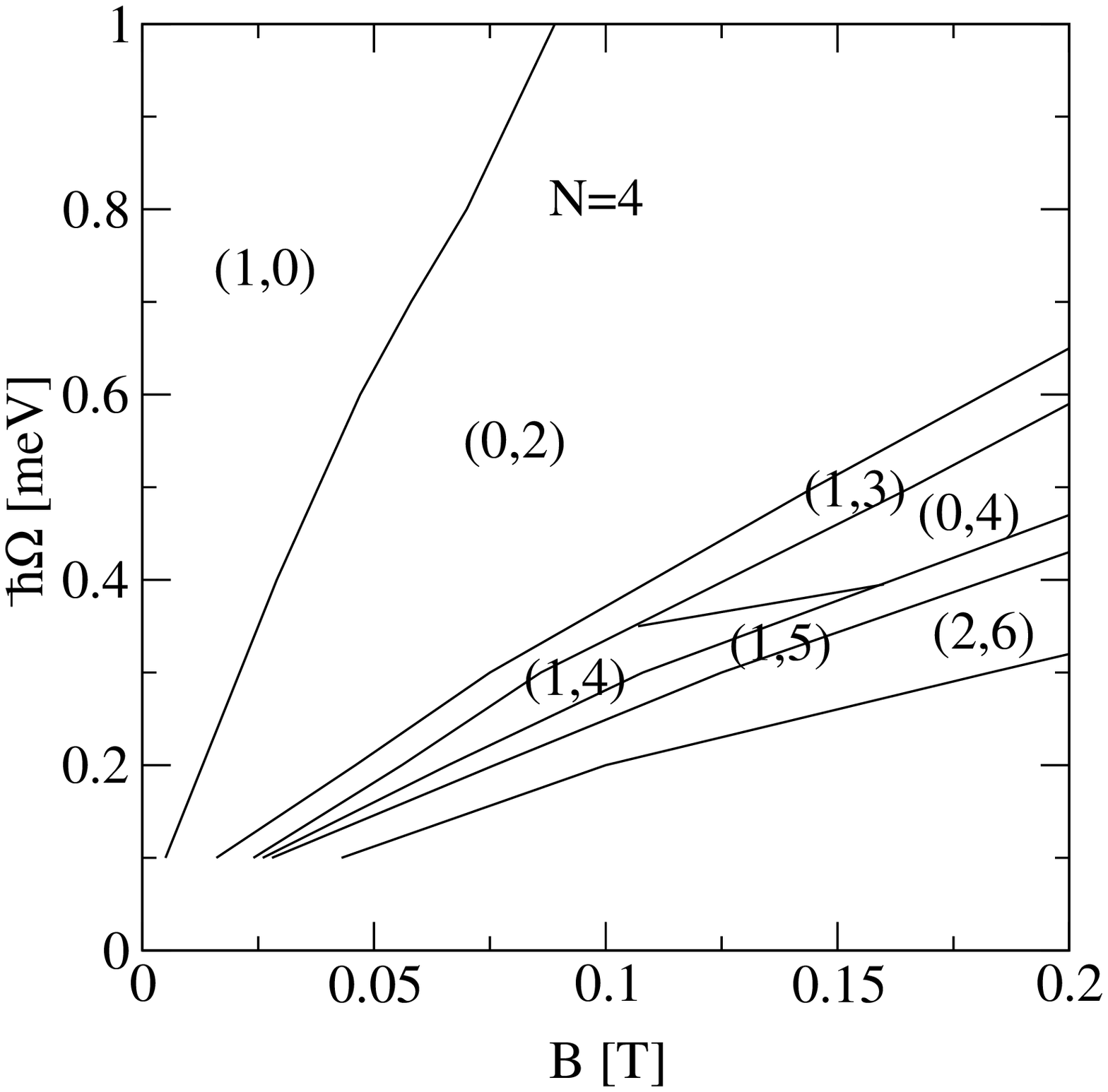}
\caption{Same as in Fig.1, but $N=4$.}
\label{fig:fig2}
\end{figure}

\begin{figure}
\center
\includegraphics*[width=2.5in,height=2.5in]{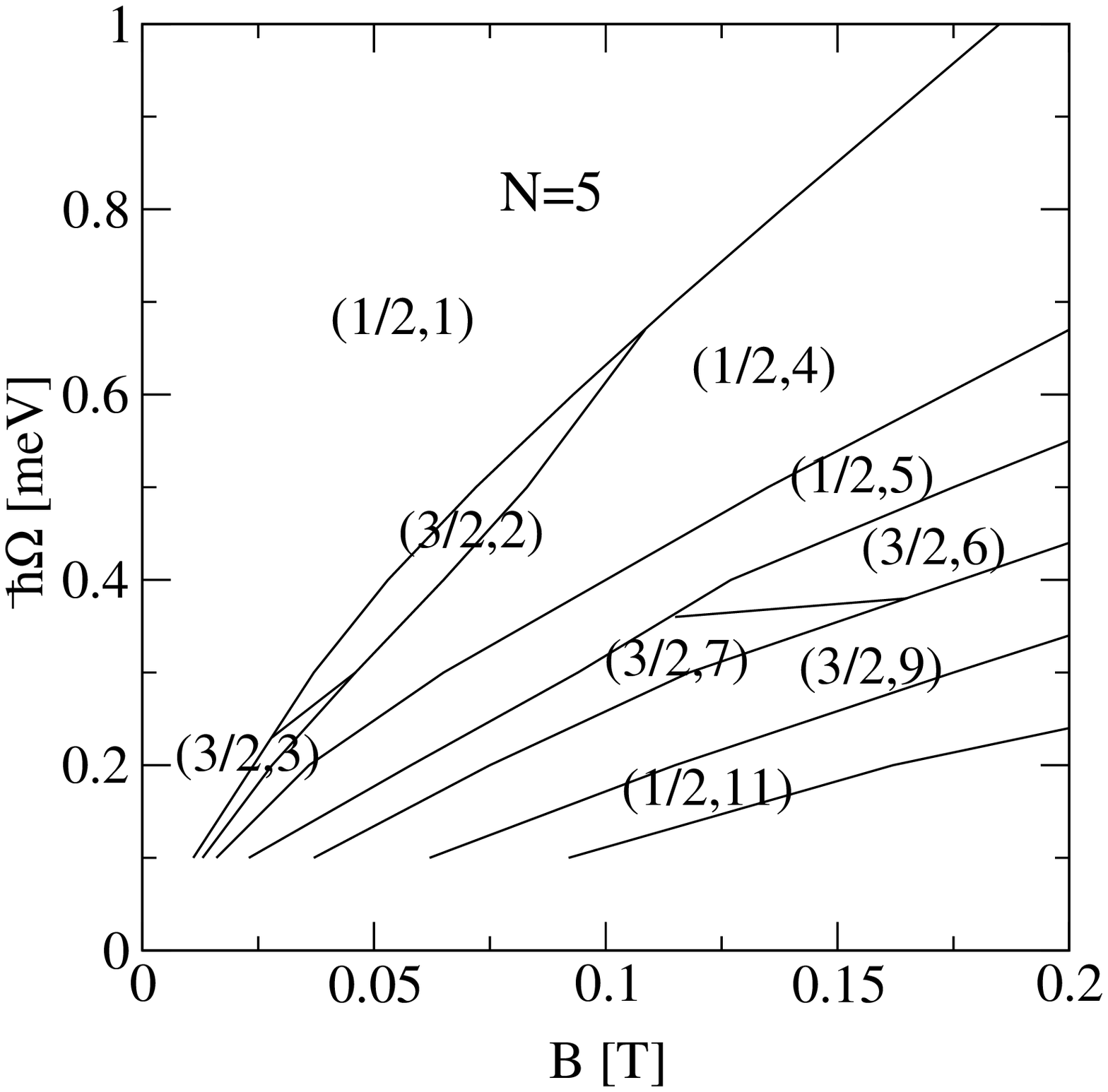}
\caption{Same as in Fig.1, but $N=5$.}
\label{fig:fig3}
\end{figure}

\begin{figure}
\center
\includegraphics*[width=2.5in,height=2.5in]{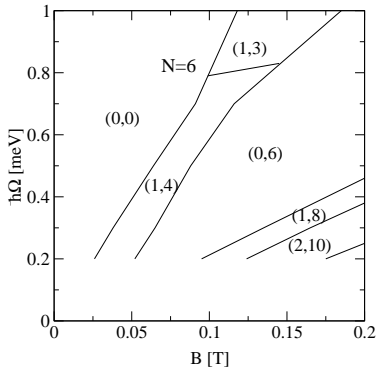}
\caption{Same as in Fig.1, but $N=6$.}
\label{fig:fig4}
\end{figure}

For numerical diagonalizaton of the Hamiltonian matrix,
only Slater determinant basis states with  energies
$E_{\alpha} \le q \hbar \omega$ are included, where  $q=20 - 25$.
Typically, we have included about 3,000-14,000 basis states.
We have checked that for these values $q$ the energy differences
between different states have converged.
For example, when  $N=6$, $B=0$ and $\hbar\Omega=0.3$meV
the energy differences  
between the ground state $(0,0)$ and one of the competing state $(2,0)$
are  0.0371, 0.0332, 0.0328 meV for q=21, 23, 25, respectively.
Results for $N=3$ and 4 are in good agreement with
the results of Mikhailov \cite{mik}. 
For example, for $\hbar\Omega=0.118565$ (corresponding to $\lambda=10$
in Ref.~[7]), we have $E_{(1/2,1)}/\hbar\Omega=17.6279 (N=3)$ and
$E_{(1,0)}/\hbar\Omega=31.4120 (N=4)$, perfectly consistent with
the numbers in the reference.
Note that when $\hbar\Omega$ gets too small more basis states need to be
included and the numerical work becomes unwieldy.  These results are not included
in the phase diagrams.

Figures~1,2,3  and 4 show phase diagrams for $N=3,4,5$, and $6$
in the weak potential and magnetic field regime.
Each phase is labeled by the total spin and angular momentum quantum
numbers $(S_z,L_z)$.
These phase diagrams are constructed schematically from the calculated
 energies of different $(S_z,L_z)$ states at various values of $(\hbar \Omega,B)$.
We took  $B=0, 0.01,....,0.19,0.2$ T for all values of $N$ shown in the figures, and took
$\hbar \Omega=$ 0.1, 0.2, ... 1.0 meV for $N=3,4,5$ while
$\hbar \Omega=$ 0.2, 0.3, 0.5, 0.7, 1.0 meV for $N=6$.
We remark on some of the qualitative aspects of these phase diagrams.
(i)Note that some the phase regions are divided into subregions separated
by horizontal phase boundaries. Also we notice that some of the phase boundaries
are  linear functions of $B$. 
(ii)These figures  show that, as the strength of the potential or  magnetic field changes 
the ground state quantum number $L_z$ changes 
{\it sensitively}.  As the magnetic field $B$ increases 
the value of $L_z$ tend to increase.  On the other hand 
as  the strength of the confinement
potential $\hbar\Omega$ increases the ground state $L_z$ tend to decrease.
(iii)Ground state energies obtained for 
$L_z=0,3,4$, and $6$ are,  respectively, 37.28, 37.58, 37.88, and 38.05 meV
(These values are for the parameters $N=6$, $B=0$,
and $\hbar \Omega= 1.0$ meV).
These results indicate that the ground states are nearly
degenerate as a result of strong electron correlation.

\begin{figure}
\center
\includegraphics[width=2.8in]{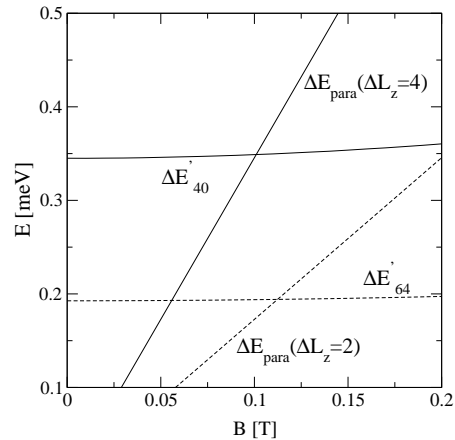}
\caption{Here we compare the differences
$E_{para}(S_{z,2},L_{z,2})-E_{para}(S_{z,1},L_{z,1})$ 
and $E^{'}(S_{z,1},L_{z,1})-E^{'}(S_{z,2},L_{z,2})$ for states sharing the same phase boundary.
Two solid lines represent the magnetic field dependence of
$\Delta E_{para}(\Delta L_z=4)=E_{para}(0,0)-E_{para}(1,4)$
and $\Delta E^{'}_{40}=E^{'}(1,4)-E^{'}(0,0)$, and the other two dotted lines represent
$\Delta E_{para}(\Delta L_z=2)=E_{para}(1,4)-E_{para}(0,6)$
and $\Delta E^{'}_{64}=E^{'}(0,6)-E^{'}(1,4)$.
When these pair of lines cross the ground state changes.
Here $N=6$, $\hbar \Omega=0.7 {\rm meV}$, and  $E_{spin}$ is ignored.
Note that the differences in $E^{'}(S_{z},L_{z})$ are almost constant   
as a function of magnetic field while the  difference in $E_{para}(S_{z},L_{z})$
are significant.} 
\label{fig:fig5}
\end{figure}
We now discuss why  some of the phase boundaries in Figures 1, 2, 3, and 4 are  linear functions of $B$.
It is useful to separate the total energy into 
\begin{eqnarray}
E_{tot}=E_{kin}+E_{diam}+E_{para}+E_{spin}+E_{int},
\end{eqnarray}
where
$E_{kin}=<\sum_{i}\epsilon^0_{n_il_i}>$ is the kinetic energy,
$E_{diam}=<\sum_{i}\frac{1}{8}m^*\omega_c^2r_i^2>$
is the diamagnetic energy, $E_{para}=-\frac{1}{2}\omega_cL_z$ is 
the paramagnetic energy, $E_{int}$ is the interaction energy, and  
$E_{spin}=-g\mu B S_z$ is the Zeeman energy.
Note that $E_{kin}\propto \Omega$.  We now compare $E_{para}$ with the rest of the ground state energy,
$E^{'}(S_z,L_z)=E_{kin}+E_{diam}+E_{int}$, ignoring $E_{spin}$, which is very small
compared to the others. 
For many states found in the phase diagrams the values of $E^{'}(S_z,L_z)$ are nearly the same.
This property reflects the delicate competition between different states, which is an indication
that the quantum dot is in the strongly correlated regime.  However, since $E_{para}$
depends significantly on $B$, we expect  ground states
to change sensitively as a function of $B$.
This is explained in detail in Figure~5.
Phase boundaries in the  
parameter space $(\hbar\Omega, B)$ is roughly determined by
the sum of kinetic and paramagnetic energies: 
Consider the phase boundary between
ground states $(S_{z,1},L_{z,1})$ and $(S_{z,2},L_{z,2})$, then, since 
the values of  $E^{'}(S_{z,i},L_{z,i})$ are nearly the same, we expect   
$b\Omega_1-\frac{1}{2}\omega_cL_{z,1}\approx b\Omega_2-\frac{1}{2}\omega_cL_{z,2}$,
which gives $\Omega_1-\Omega_2\approx\frac{\omega_c}{2b}(L_{z,1}-L_{z,2})$
(here $b $ is the proportionality
factor in the kinetic energy).
This result indicates that some phase boundaries are approximately a linear function of $B$.
Also the slope of such a phase boundary is proportional
to the difference in the angular momentum
$L_{z,1}-L_{z,2}$.
These results are consistent with many phase boundaries in 
Figures~1, 2, 3, and 4, although there are few exceptions. 

\begin{figure}
\center
\includegraphics[width=8cm]{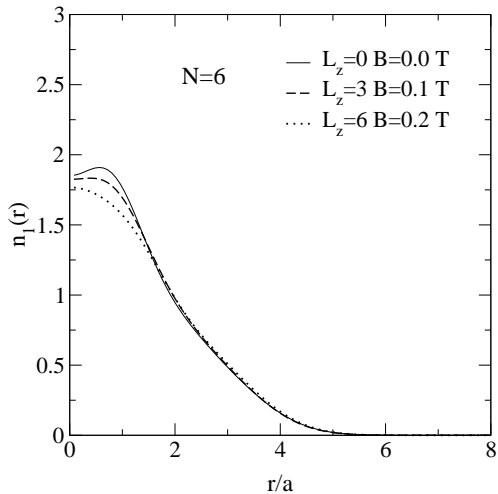}
\caption{Radial density profiles of several different groundstates with $S_z=0$.  Here 
$\hbar \Omega= 0.7 meV$ and
$N=6$.   
For weak magnetic fields, radius of dot is roughly $5 a$,
where $a=\sqrt{\hbar/2m\omega}$.    We notice that the profiles 
deviate slightly from each other.  From these profiles we can estimate 
the density of 2D 
electron gas which these dots are made of to be about $9\times 10^9 cm^{-2}$.}
\label{fig:fig6}
\end{figure}
\begin{figure}
\center
\includegraphics[width=8cm]{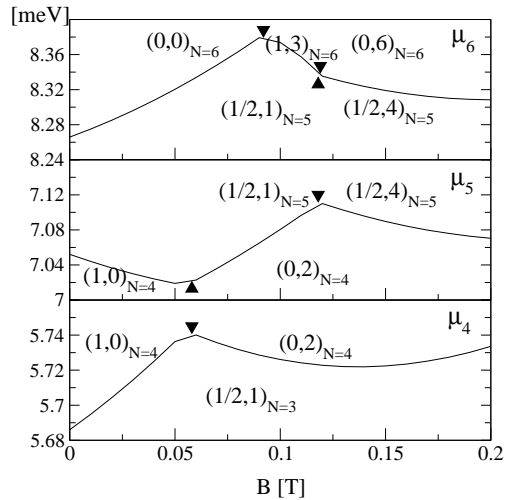}
\caption{
Plot of the chemical potential $\mu_N \equiv E_N - E_{N-1}$            
as a function of the magnetic field at $\hbar\Omega=0.7$ meV for $N=4,5$,       
and 6.  
The positions of magnetic fields, where a ground state level crossing in the 
$N-1$ and $N$ electron systems takes place, are 
indicated by $\bigtriangleup$ and $\bigtriangledown$.                                                   
The quantum states of the dots are denoted by $(S_z,L_z)$
(The other cusp-like structures are artifacts of discrete data points).
}
\label{fig:fig7}
\end{figure}
We now discuss some issues of experimental relevance.
First we estimate the density of 2D electron gas which these dots can be made of.
Figure~\ref{fig:fig6} shows the radial electron density profile of dots for
$N=6$ when $\hbar \Omega= 0.7 meV$. The radius of the dot is
roughly $5 a$ for weak fields considered in this paper.  We have verified by integrating the density profile
that the total number of electrons is $N$.   For $m^{*}=0.067m$, where $m$ is the bare electron mass,
the average electron density is approximately
$9\times 10^9 cm^{-2}$.
We suggest that our obtained  phase diagrams may be explored experimentally
by measuring 
the positions of cusps in the magnetic field dependence of 
chemical potential $\mu_N \equiv E_N - E_{N-1}$:
Since ground state level crossings in the $N-1$ and $N$ electron systems lead respectively to
positive and negative cusps, phase boundaries will show up as cusps in the chemical potential.  
We plot the magnetic field dependence of the chemical potentials $\mu_4$,
$\mu_5$, and $\mu_6$,
in Figure~\ref{fig:fig7}.   
We observe several
cusps in Fig.~\ref{fig:fig7}, where $\hbar\Omega=0.7$ meV is used.
From our phase diagrams we can conclude that for smaller
values of $\hbar\Omega$ 
many more cusps will show up in the magnetic field dependence of the chemical potential.
Conductance peak spacings, given as the difference between  the gate voltages, $V_g^{N+1}-V_g^{N}$,
are related to the positions of the cusps: i.e.
$e (V_g^{N+1}-V_g^{N}) \propto (\mu_{N+1}-\mu_{N}).$
The other issue concerns that in real dots there may be some 
deviations from the perfect circular symmetry of
the harmonic potential, and consequently degenerate single particle states of the harmonic potential
at $B=0$ may be absent.
However, even in a harmonic potential electron-electron interactions will lift
this degeneracy.
Furthermore, the main conclusion of our work that the ground states change
sensitively with $B$ or $\hbar\Omega$ is not expected to be different
since strong electron correlation is responsible
for the sensitivity of ground states  to the magnetic field and the strength of the
confinement potential.  

In this paper we explored  quantum dots in a strongly
correlated regime $R \gg 1$.
Our exact diagonalization results show that
the ground states  of such  quantum dots 
are  rather sensitive to the magnetic field and  strength of the
confinement potential.  We have predicted rich phase diagrams.
Experimentally these phase diagrams
may be explored by measuring the sensitive magnetic field dependence of the energy to add one electron 
to a dot, or by measuring the positions of conductance peak oscillations.

S.R.E.Y was supported by KOSEF through the Quantum-functional
Semiconductor Research Center
at Dongguk University and by grant R01-1999-00018-0
from interdisciplinary Research program of KOSEF.

\end{document}